# Neutron Scattering Studies of the Ferroelectric Distortion and Spin Dynamics in the Type-1 Multiferroic Perovskite $Sr_{0.56}Ba_{0.44}MnO_3$

Daniel K. Pratt,[1] Jeffrey W. Lynn,[1]* James Mais,[2] Omar Chmaissem,[2,3] Dennis E. Brown,[2] Stanislaw Kolesnik,[2] and Bogdan Dabrowski[2]

[1]*NIST Center for Neutron Research, National Institute of Standards and Technology, Gaithersburg, MD 20899-6102*
[2]*Department of Physics, Northern Illinois University, DeKalb, IL 60115*
[3]*Materials Science Division, Argonne National Laboratory, Argonne, IL 60439*

The magnetic order, spin dynamics, and crystal structure of the multiferroic $Sr_{0.56}Ba_{0.44}MnO_3$ have been investigated using neutron and x-ray scattering. Ferroelectricity develops at $T_C=305$ K with a polarization of 4.2 µC/cm$^2$ associated with the displacements of the Mn ions, while the $Mn^{4+}$ spins order below $T_N \approx 200$ K into a simple G-type commensurate magnetic structure. Below $T_N$ the ferroelectric order decreases dramatically demonstrating that the two order parameters are strongly coupled. The ground state spin dynamics are characterized by a spin gap of 4.6(5) meV and the magnon density of states peaking at 43 meV. Detailed spin wave simulations with a gap and isotropic exchange of $J=4.8(2)$ meV describe the excitation spectrum well. Above $T_N$ strong spin correlations coexist with robust ferroelectric order.

PACS: 75.85.+t; 75.30.Ds; 78.70.Nx; 77.80.B-;

*Corresponding author: Jeffrey.Lynn@nist.gov

Multiferroics are systems that exhibit both ferroelectric order and magnetic order, and are a topic of current interest both to understand how these two disparate order parameters interact and because of the intriguing possibility of controlling the magnetic properties electronically and vice versa [1-3]. Type-II or 'improper' multiferroics have been a focus of fundamental studies as they exhibit strong magnetic-ferroelectric coupling, often with the magnetic order that develops at low temperatures breaking the structural centrosymmetry and enabling ferroelectric order to develop. However, the ferroelectric order parameter is typically two to three orders-of-magnitude smaller than robust ferroelectrics such as the prototypical $BaTiO_3$ used in contemporary applications. The strong ferroelectricity in materials such as $BaTiO_3$ originates from the bonding of the occupied oxygen *p* orbitals to the empty Ti *d* orbitals, and with the *d* orbitals empty this excludes the possibility of magnetic order. Thus type-I or 'proper' multiferroics with robust ferroelectric order are not only quite rare but also typically with disparate ordering temperatures and weak coupling between the order parameters. Therefore for multiferroics to be useful an alternative mechanism for ferroelectricity is highly desirable, where higher order temperatures and strong coupling might be realized. One such alternative mechanism for magnetic—ferroelectric coupling was pointed out for a class of perovskite manganites such as $(R,Ca)MnO_3$ (R=lanthanide) via the charge and/or orbital ordering that can occur in these systems [4], which is a class of materials perhaps more familiar as colossal magnetoresistive materials at lower doping [5]. The advantage of the perovskite manganites is that these are type-I multiferroics that can exhibit a large electric polarization. First principles calculations have been carried out for the $BaMnO_3$ [6] and $SrMnO_3$ [7] end members of the $Sr_{1-x}Ba_xMnO_3$ series, as well as for $CaMnO_3$ [8], which predict the existence of a ferroelectric ground state for the Ba system and ferroelectricity for the Sr material if it is strained, with an electric polarization comparable to BaTiO$_3$ (tens of µC/cm$^2$). Work on lightly substituted Sr$_{1-x}$Ba$_x$MnO$_3$ up to a maximum of x≈0.2 with conventional fabrication techniques did not reveal ferroelectricity [9], while more recent work to higher concentrations identified the expected soft phonon that should trigger the ferroelectric order at higher x [10]. Indeed with further increase of Ba robust ferroelectricity has been recently observed [11]. Here we present neutron scattering results for the spin dynamical properties of the type-I multiferroic perovskite Sr$_{0.56}$Ba$_{0.44}$MnO$_3$ and the coupling of the magnetic order parameter to the ferroelectricity. The system becomes ferroelectric above 400 K depending on composition as shown in Fig. 1(a), with a ferroelectric transition temperature exceeding and polarization comparable to BaTiO$_3$, and develops antiferromagnetic order at ≈200 K. The ferroelectric order is associated with the displacements of the Mn ions while the magnetism originates with the Mn magnetic moments, guaranteeing that these two order parameters are strongly coupled.

For this work polycrystalline samples of perovskite Sr$_{1-x}$Ba$_x$MnO$_3$ were synthesized using the two-step solid state method described in references [9,11,12], which is required to avoid the more stable hexagonal polymorph and remain in the pseudocubic regime where the ferroelectric order exists. First the hexagonal precursor material was synthesized in flowing H$_2$/Ar gas at high temperatures near 1300 C to obtain single-phase oxygen-reduced perovskite samples. In the second step, the reduced samples were annealed in oxygen at 350 C resulting in the full oxygen stoichiometry of 3.00 as verified by precise thermogravimetric measurements. The phase diagram for this system obtained from our polycrystalline samples and the x-0.5 singe crystal of Sakai, *et al.* [11] is shown in Fig. 1(a). Just above x ≈ 0.4 ferroelectric order develops with T$_C$ increasing rapidly with x to above 400 K. For the same samples the magnetic ordering develops around 200 K, below which the ferroelectric order is partially or fully suppressed [11].

High resolution x-ray diffraction measurements were taken on the 11-BM-B (λ=0.413283 Å) and 11-ID-C (λ=0.10801 Å) beam lines at the Advanced Photon Source, Argonne National Laboratory [13]. Diffraction patterns were collected between 120 and 440 K on heating using a commercially available cryostream system. For the neutron scattering work presented here we focus on the composition with x=0.44. Elastic and inelastic neutron measurements of the magnetic order parameter and spin dynamics were taken on the BT-7 triple axis spectrometer at the NIST Center for Neutron Research [14], where pyrolytic graphite monochromator, analyzer, and filters were employed with a fixed final energy of 14.7 meV. Collimations of 60'-80'-80'-120' with flat analyzer and 60'-80'-80R'-180' with horizontal focusing for the analyzer were used for the inelastic measurements, while for the powder diffraction measurements the position sensitive detector option on BT-7 was employed to collect diffraction data and the magnetic order parameter. The SPINS cold triple axis spectrometer was utilized to provide higher energy resolution in measuring the spin gap more accurately. Pyrolytic graphite monochromator and analyzer were used with collimations of guide-80'-80'-open and a final energy of 5.0 meV, with a cold polycrystalline Be filter in the scattered beam to suppress higher order contaminations. For these neutron measurements the sample consisted of ~4 grams of Sr$_{0.56}$Ba$_{0.44}$MnO$_3$ powder that was loaded into an aluminum powder can, sealed with helium exchange gas. and placed in a closed cycle refrigerator with a temperature range of 2.5 K to 325 K.

The x-ray diffraction data for the {2,0,0} structural peak taken on the x=0.44 composition are shown in Fig. 1(b) as a function of temperature. At high temperatures the crystal structure is cubic, space group symmetry of $Pm\bar{3}m$ (a = 3.8693(1) Å at 440 K), but below 305 K the peak splits into {2,0,0} and {0,0,2} peaks, indicating a sluggish transition to tetragonal symmetry ($a$=$b$=3.85 Å, $c$=3.88 Å; $c/a > 1$) where ferroelectricity develops. A detailed crystallographic analysis of the $Mn^{4+}$ displacements reveals a ferroelectric order parameter of 4.2 μC/cm². This is in very good agreement with the value of 4.5 μC/cm² reported in ref. 11 for such a heavily twinned crystal, which would imply an intrinsic polarization of ≈13 μC/cm² as discussed there. The ordering temperature and polarization increase with increasing x as indicated in Fig. 1(a). These values are comparable to both the measured and calculated values of the ordering temperature and polarization for $BaTiO_3$, with a polarization of 27 μC/cm² and a transition temperature of 393 K [15, 16].

Fig. 1(b) also shows a dramatic decrease in the tetragonal splitting when the system orders antiferromagnetically at around 200 K. One interesting aspect of this transition is that susceptibility measurements up to applied fields of 7 T indicate that the magnetic ordering temperature is insensitive to field. The magnetic structure is the very simple G-type, where nearest neighbor spins are antiparallel. The structure at low temperature is basically cubic again, with only a slight tetragonal distortion remaining for this composition, so that the ferroelectric order parameter is greatly reduced. The crystal structure analysis reveals that the O-$Mn^{4+}$-O bond angle increases to very close to 180º as the antiferromagnetic order develops, which maximizes the antiferromagnetic exchange interaction as discussed in references [9-11]. We identify this behavior as the origin of the strong coupling between the ferroelectric and magnetic order parameters.

To explore the spin dynamics, inelastic neutron scattering measurements were carried out on BT-7. Strong magnetic scattering was observed emanating from the {1/2,1/2,1/2} magnetic Bragg peak position located at $|\mathbf{Q}|= |\mathbf{Q}_{AFM}|= 1.42$ Å$^{-1}$. Fig. 2(a) shows a contour plot of the intensity of the scattering at 2.5 K, where the width of the scattering increases with increasing energy as the spin waves disperse. There is a weak maximum in the intensity at ≈7 meV, and the scattering decreases at lower energies indicating that there is a significant spin gap in the system.

In order to quantitatively analyze the inelastic neutron scattering measurements in $Sr_{0.56}Ba_{0.44}MnO_3$, a simple Heisenberg model was employed for the analysis. Because of the polycrystalline nature of the sample and the near-cubic crystal structure it would be difficult to accurately extract any anisotropy information in the exchange parameters, so for our analysis we assumed an isotropic coupling ($J_a$=$J_b$=$J_c$). Therefore, the Hamiltonian used to describe the system is given by

$$H = -J \sum_{i,j} S_i \cdot S_j - g\mu_B H_a \sum_i \sigma_i S_i , \qquad (1)$$

where $J$ represents the isotropic magnetic exchange between Mn atoms. The uniaxial single-ion anisotropy (which results in the spin gap $\Delta_{AFM}$) is represented by an anisotropy field $H_a$ that acts on spin $S_i$ and points along the spin direction given by $\sigma_i = \pm 1$. The standard linear spin wave approximation was employed to calculate the dispersion $\omega_n(\mathbf{q})$ and $S_{mag}(\mathbf{Q},\omega)$ and then perform the powder averaging to fit to the neutron data and determine the magnetic exchange interaction $J$ [17,18].

Figure 2(b) shows a contour plot of the resulting scattering intensity calculation after convolution with the instrumental energy resolution. These calculations can be directly compared to the BT7 measurements in Fig. 2(a), where we see that overall there is quite good agreement. Figure 2(c) shows constant energy scans at four different values through the spin wave scattering at the Bragg peak location ($|\mathbf{Q}_{AFM}|= 1.42$ Å$^{-1}$) for direct comparison of theory and experiment. The solid curves are the calculated $|\mathbf{Q}|$ scans, which agree quite well with the data. However, there is some structure in the data that is not reproduced by the numerical calculations, which may indicate the effect of further neighbor exchange interactions.

The low energy spin excitations provide a good first estimate of the exchange and spin gap, and Fig. 3(a) displays the overall magnon excitations expected from the polycrystalline sample while the calculated single crystal dispersion curves for two high-symmetry directions are shown in Fig. 3(b). At high energies a peak in the magnetic scattering is expected that is associated with the top of the magnon density-of-states as the spin wave dispersion relations flatten on approaching the Brillouin zone boundaries. The strongest scattering at Q < 1 is difficult to access with thermal neutrons as one needs to use very high incident energy neutrons to conserve momentum and energy in the scattering process. Hence we carried out a constant-Q measurement for $|\mathbf{Q}| = 4.2$ Å$^{-1} \approx 3|\mathbf{Q}_{AFM}|$, scanning in energy from 25 to 55 meV. The observed scattering is shown in Fig. 3(c) where the peak occurs at 43(1) meV and clearly reveals the top of the magnon band. For our system which exhibits G-type antiferromagnetic order the top of the band is given by the equation zS$J$ [18] where z = 6 is the number of nearest neighbors, S = 3/2 is the spin of the Mn$^{4+}$ ion and $J$ is the exchange interaction. The observed peak is in excellent agreement with the calculations (solid curve). Therefore, our exchange interaction is $|J| = 4.8(2)$, which is comparable to the value observed in the similar LaFeO$_3$ compound which also has a G-type AFM ground state [18].

Finally, to improve the determination of the gap in the spin wave excitation spectrum we carried out constant-Q and constant-E scans at low temperature on the SPINS cold neutron triple-axis instrument. Fig. 3(d) shows the data, which reveal a sharp increase in the scattering above ≈4 meV. Constant-E scans (not shown) reveal no observable magnetic scattering below 4 meV, indicating a clean spin wave gap in the system. The solid curve is a fit to a Heaviside function convoluted with the experimental resolution, which provides an excellent description of the data and reveals a spin gap $\Delta_{AFM} = 4.6(5)$ meV. Thus our triple axis measurements have revealed all parameters in the Heisenberg model and the ground state spin excitation spectrum has been completely determined.

The scattering is resolution limited experimentally at low temperature as confirmed by tightening the wave vector resolution with a corresponding reduction in the observed width. With increasing temperature initially the intensities of the spin wave excitations increase due to the Bose-Einstein population factor for bosons. Figure 4 shows the temperature dependence of the scattering for an energy transfer of 5 meV, just above the low temperature spin gap in the excitation spectrum. One sees a clear increase in the intensity of the spin wave scattering up to 175 K, along with a modest broadening of the scattering at these elevated temperatures. We also see an increase in the 'background', which originates primarily from multi-magnon scattering as the spin excitations become heavily populated. In addition, we find that the spin gap itself decreases (not shown) and approaches zero as the transition is approached from below. The data in Fig. 4 at 200 K, just above the transition temperature, reveal that the scattering has broadened quite substantially when long range order is lost. With further increase of temperature the data confirm the expected decrease in the spin correlations (width of scattering increases in wave vector), but there still are substantial spin-spin correlations in the paramagnetic state up to at least 1.5 $T_N$.

The present results have fully determined the magnetic properties in terms of the nearest-neighbor exchange $J$=4.8(2) meV and spin gap $\Delta_{AFM} = 4.6(5)$ for these very interesting manganites that possess robust ferroelectric order that develops above room temperature. For this particular system the magnetic order also occurs at relatively high temperatures compared to type-II multiferroics, although still below room temperature. In many multiferroics the magnetic and ferroelectric domains are coupled together, which facilitates control of the magnetic properties with an electric field and vice versa [19,20]. In the present $Sr_{1-x}Ba_xMnO_3$ system the magnetism and ferroelectricity are very strongly coupled as evidenced by the dramatic change in the electric polarization when the material orders magnetically. We speculate that in this paramagnetic-ferroelectric regime magnetic fields can induce a substantial net magnetization that couple the two types of order, and through the magnetoelectric coupling an electric field should induce a magnetization that could prove to a very useful property even though there is no long range magnetic order [21,22]. Of course, it would be beneficial to be able to tune the exchange and concomitant magnetic ordering temperature near or above room temperature where the strongest coupling would be available [22-27]. Possible pathways to achieve this would be chemical substitution or perhaps strain for thin film applications.

## Acknowledgements

The authors would like to acknowledge R. J. McQueeney for the use of the spin dynamics code as well as advice employing the code in our study, and Steven Disseler, Ronald Cappelletti, and William Ratcliff for their assistance. DKP would like to acknowledge support from the National Research Council NIST postdoctoral associateship program. OC acknowledges support by the U.S. Department of Energy, Office of Science, Materials Sciences and Engineering Division. Use of the Advanced Photon Source at Argonne National Laboratory was supported by the U. S. Department of Energy, Office of Science, Office of Basic Energy Sciences, under Contract No. DE-AC02-06CH11357.

**References:**
[1]  S-W. Cheong and M. Mostovoy, Nature Mater. **6**, 13 (2007).
[2]  R. Ramesh and N. A. Spaldin, Nature Mater. **6**, 21 (2007).
[3]  D. Khomskii, Physics **2**, 20 (2009).
[4]  D. V. Efremov, J. van den Brink, and D. I. Khomskii, Nature Mater. **3**, 853 (2004).
[5]  See, for example, S.-W. Cheong and C. H. Chen, in *Colossal Magnetoresistance, Charge Ordering and Related Properties of Manganese Oxides*, edited by B. Raveau and C. N. R. Rao (World Scientific, Singapore, 1998); E. Dagotto, *Nanoscale Phase Separation and Colossal Magnetoresistance* (Springer, Berlin, 2003).
[6]  James M. Rondinelli, Aaron S. Eidelson, and Nicola A. Spaldin, Phys. Rev. B **79**, 205119 (2009).
[7]  Jun Hee Lee and Karin M. Rabe, Phys. Rev. Lett. **104**, 207204 (2010).
[8]  S. Bhattacharjee, Eric Bousquet, and Philippe Ghosez, *Phys. Rev. Lett*. **102**, 117602 (2009).
[9] O. Chmaissem, B. Dabrowski, S. Kolesnik, J. Mais, D. E. Brown, R. Kruk, P. Prior, B. Pyles, and J. D. Jorgensen, Phys. Rev. B**64**, 134412 (2001).
[10]  H. Sakai, J. Fujioka, T. Fukuda, M. S. Bahramy, D. Okuyama, R. Arita, T. Arima, A. Q. R. Baron, Y. Taguchi, and Y. Tokura, *Phys. Rev. B* **82**, 104407(R) (2010).
[11]  H. Sakai, J. Fujioka, T. Fukuda, D. Okuyama, D. Hashizume, F. Kagawa, H. Nakao, Y. Murakami, T. Arima, A. Q. R. Baron, Y. Taguchi, and Y. Tokura, *Phys. Rev. Lett.* **107**, 137601 (2011).
[12] B. Dabrowski, O. Chmaissem, M. Mais, S. Kolesnik, J.D. Jorgensen, and S. Short, J. Solid State Chem. **170**, 154 (2003).
[13] J. Wang, B. H. Toby, P. L. Lee, L. Ribaud, S. M. Antao, C. Kurtz, M. Ramanathan, D. B. Von Dreele, M. A. Beno, Rev. Sci. Instrum. **79**, 085105 (2008).

[14] J. W. Lynn, Y. Chen, S. Chang, Y. Zhao, S. Chi, W. Ratcliff, B. G. Ueland, and R. W. Erwin, Journal of Research of NIST **117**, 61 (2012).

[15] See, for example, M. E. Lines and A. M. Glass: *Principles and Applications of Ferroelectrics and Related Materials* (Oxford University Press, Oxford, 1977).

[16] W. Wong, D. Vanderbilt, and K. M. Rabe, Phys. Rev. Lett. **73**, 1861 (1994).

[17] R. J. McQueeney, *Phonon* and *Spinwaves* software suite (2008).

[18] R. J. McQueeney, J-Q. Yan, S. Chang, and J. Ma, Phys. Rev. B **78**, 184417 (2008).

[19] B. G. Ueland, J. W. Lynn, M. Laver, Y. J. Choi, and S.W. Cheong, Phys. Rev. Lett. **104**, 147204 (2010).

[20] Yusuke Tokunaga, Nobuo Furukawa, Hideaki Sakai, Yasujiro Taguchi, Taka-hisa Arima, and Yoshinori Tokura, Nature Materials 8, 558 (2009).

[21] Electric-field induced ferromagnetic phase in paraelectric antiferromagnets, Maya D. Glinchuk, Eugene A. Eliseev, Yijia Gu, Long-Qing Chen, Venkatraman Gopalan, and Anna N. Morozovska, Phys. Rev. B **89**, 014112 (2014).

[22] G. Srinivasan, Annu. Rev. Mater. Res. **40**, 153 (2010).

[23] R. O. Cherifi, V. Ivanovskaya, L. C. Phillips, A. Zobelli, I. C. Infante, E. Jacquet, V. Garcia, S. Fusil, P. R. Briddon, N. Guiblin, A. Mougin, A. A. Unal, F. Kronast, S. Valencia, B. Dkhil, A. Barthelemy, and M. Bibes, Nat. Mater. **13**, 345 (2014).

[24] J. Wang, J. B. Neaton, H. Zheng, V. Nagarajan, S. B. Ogale, B. Liu, D. Viehland, V. Vaithyanathan, D. G. Schlom, U. V. Waghmare, N. A. Spaldin, K. M. Rabe, M. Wuttig, R. Ramesh, Science **299**, 1719 (2003).

[25] Carlos Antonio Fernandes Vaz and Urs Staub, J. Materials Chemistry C (in press).

[26] Electric field control of magnetic properties and magneto-transport in composite multiferroics, O. G. Udalov, N. M. Chtchelkatchev, and I. S. Beloborodov, http://arxiv.org/abs/1404.6671v1

[27] M. Bibes and A. Barthelemy, Nat. Mater. **7**, 425 (2008).

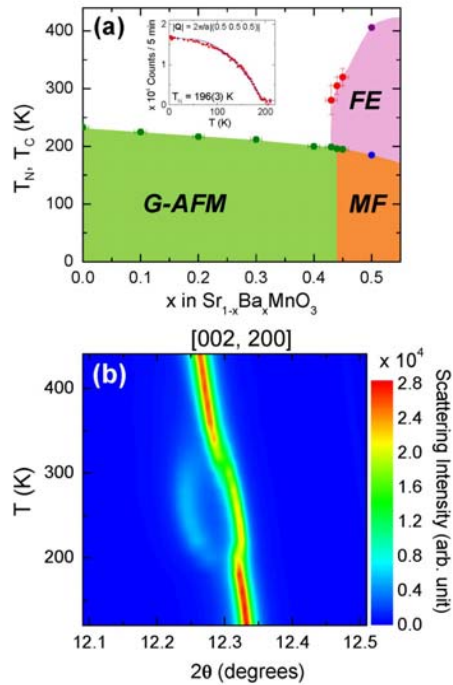

Figure 1 (color online). (a) Phase diagram for $Sr_{1-x}Ba_xMnO_3$ showing the concentration region where ferroelectric (FE) and multiferroic (MF) behavior develop. The point for x=1/2 is from Sakai, *et al.* [11]. The present multiferroic sample studied in this work is $Sr_{0.56}Ba_{0.44}MnO_3$. The inset shows the smooth development of the integrated intensity of the {1/2,1/2,1/2} antiferromagnetic peak of this G-type magnetic structure obtained on BT7. The curve is a simple mean field order parameter fit of the square of the magnetic order parameter to estimate the antiferromagnetic transition $T_N$ of 196(3) K. (b) X-ray diffraction data taken at the Advanced Photon Source demonstrating the crystallographic distortion associated with the ferroelectric transition at $T_C$ = 305 K. The size of the ferroelectric distortion indicates a maximum polarization of 13 $\mu C/cm^2$ at this composition. Note that the distortion is dramatically reduced as the magnetic order develops, demonstrating that the two order parameters are strongly coupled.

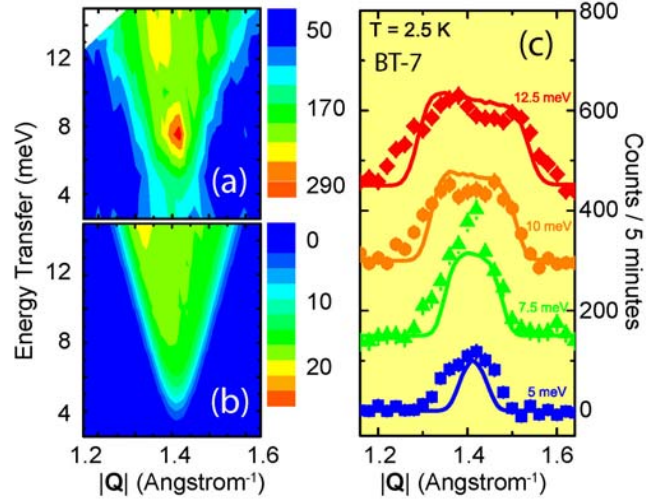

Figure 2 (color online). (a) Contour plot of the BT-7 experimental data for the ground state spin dynamics at low energies, which can be directly compared with the calculated magnon density-of-states shown in (b). (c) constant energy cuts through the low temperature data (points). Data have been offset by 150 counts for each energy for clarity. The solid curves through the data are the calculated results from the model. Uncertainties where indicated are statistical in origin and represent one standard deviation.

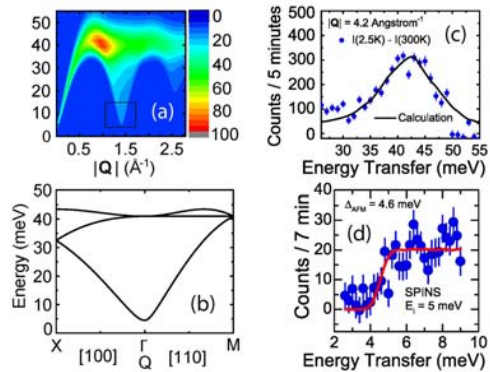

Figure 3 (color online). (a) Calculated contour map of the overall polycrystalline spin wave scattering, convoluted with the instrumental resolution, using the final parameters of J = 4.8 meV and $\Delta_{AFM}$ = 4.6 meV. The wave vector and energy ranges shown in Fig. 2(a,b) are indicated by the bounded box. (b) Calculated spin wave dispersion expected for a single crystal (c) Inelastic neutron scattering measurement showing the top of the band of magnetic excitations at 43(1) meV. The solid curve shows the fit to the Heisenberg model described in the text, powder averaged and convoluted with the experimental resolution. (d) High resolution inelastic measurement using the cold triple axis spectrometer SPINS to determine the ground state spin wave gap of $\Delta_{AFM}$ = 4.6 meV, which was determined by fitting the observed scattering intensity to a Heaviside function convoluted with the instrumental resolution (solid (red) curve).

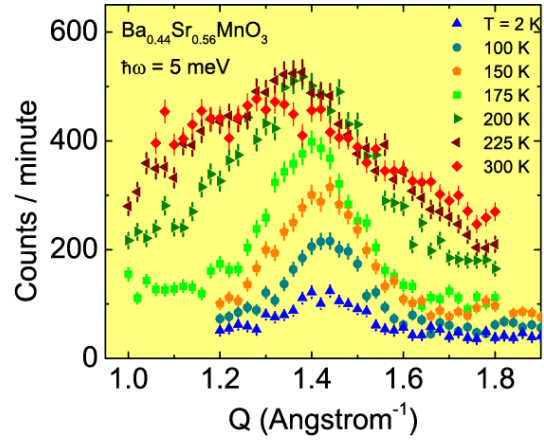

Figure 4 (color online). Temperature dependence of the scattering at an energy transfer of 5 meV. The scattering is resolution limited at low temperature. With increasing temperature the intensity increases due to the Bose-Einstein population factor for bosons. We also observe that the spin gap decreases and approaches zero as the transition is approached from below, while the wave vector width broadens in approaching and exceeding $T_N$. In the paramagnetic state the data reveal that there are substantial spin-spin correlations up to at least 1.5 $T_N$.